\documentclass[12pt]{article}
\usepackage{epsfig}
\usepackage{amsmath}
\usepackage{amssymb}

\begin{document}

\title{Study  of pure annihilation type  decays $B  \to D_s^{*} K$
}

\author{Ying Li\footnote{e-mail: liying@mail.ihep.ac.cn}\\
{\it \small    Institute of High Energy Physics,
CAS, P.O.Box 918(4), Beijing 100039, China; }\\
{\it \small   Physics Department, NanJing Normal University,
JiangSu 277600,China}\\
and  \\
Cai-Dian L\"u\footnote{e-mail: lucd@ihep.ac.cn} \\
   {\it \small   CCAST (World Laboratory), P.O. Box 8730,
   Beijing 100080, China;}\\
{\it \small    Institute of High Energy Physics,
CAS, P.O.Box 918(4) }
{\it \small Beijing 100039, China}\footnote{Mailing address.}}

\maketitle

\begin{picture}(0,0)(-320,-320)
 \put(0,32){}
 \put(0,15){BIHEP-TH-2003-13}
 \put(0,0){}
\end{picture}

\begin{abstract}
In this work, we calculate the rare decays $B^0  \to D_s^{*-} K^+$
and $B^+ \to D_s^{*+} \overline{K}^0$ in  perturbative QCD
approach with Sudakov resummation. We give the branching ratio of $10^{-5}$ for
$B^0 \to D_s^{*-}K^+$,  which
will be tested soon in $B$ factories.
 The decay $B^+ \to D_s^{*+} \overline{K}^0$ has a very small
 branching ratio at ${\cal O}( 10^{-8})$, due to the suppression
 from CKM matrix elements $|V_{ub}^* V_{cd}|$. It may be sensitive
to new physics contributions.
\end{abstract}

\newpage
\section{Introduction}

  Perturbative QCD (PQCD) method for B decays has been developed
for some years \cite{pqcd}.  It is   successfully
 applied to exclusive $B$ meson decays recently,
such as $B \to \pi\pi$ \cite{luy}, $B \to K \pi$ \cite{kls}, $B \to \pi\rho$ \cite{ly}
and other channels \cite{PQCD}.

  Very recently, pure annihilation type B decays are discussed in the PQCD approach,
  such as
 $B^0 \to D_s^-  K^+ $ and $B^+ \to D_s^+
\overline{K}^0$ \cite{lu:bdsk,ch} decays and $B^+ \to D_s^+ \phi$ decay \cite{lu}.
It is found that the annihilation type decay
 $B^0  \to D_s^{-} K^+$ has a sizable branching ratio of  $10^{-5}$,
 which has already been measured by experiments \cite{lu:bdsk}.
In this paper, we will continue to compute
the branching ratios of similar  decays $B^0  \to D_s^{*-} K^+$ and $B^+ \to D_s^{*+}
\overline{K}^0$.

Because the four valence quarks in the final states $D_s^*$ and
$K(\overline K)$ are different from the ones in the $B$ meson,
the rare decays $B \to D_s^{*} K$ are pure annihilation type
decays.  In the usual factorization approach (FA) \cite{bsw}, this decay
picture is described as $B$ meson annihilating into vacuum and
$D_s^*$ and $K$ meson produced from vacuum afterwards. To calculate
this decay in the FA, one needs the $D_s^* \to K$ form factor at
very large timelike momentum transfer ${\cal O} (M_B)$. However
the form factor at such a large momentum transfer is not known in
FA.
The annihilation amplitude is a phenomenological parameter in QCD
factorization approach (QCDF) \cite{bbns}, and the QCDF calculation
of these decays is also unreliable.

In this paper, we will calculate these decays in PQCD approach.
Similar to the $B \to D_S K$ decays,
the $W$ boson exchange induce the four quark operator
 $\bar{b}d \to \bar{c}u $ or $\bar b
u \to \bar{d}c$, and the $\bar{s}s$ quarks included in $D_s^* K$
are produced from a gluon. This gluon  attaches to any one of the
quarks participating in the four quark operator.
In the rest frame of $B$ meson, the produced
$s$ or $\bar{s}$ quark included in $D_s^* K$ final states has
$\mathcal{O}(M_B/4)$ momentum, and the gluon producing them
has $q^2 \sim \mathcal{O}(M_B^2/4)$. This is a hard gluon, so we
can perturbatively treat the process where the four-quark operator
exchanges a hard gluon with $s \bar s$ quark pair. It is a
perturbative six quark interaction now.

In PQCD, the decay amplitude is separated  into soft ($\Phi$),
hard ($H$), and harder ($C$) dynamics characterized by different
scales. It is conceptually written as the convolution,
\begin{equation}
 \mbox{Amplitude}
\sim \int\!\! d^4k_1 d^4k_2 d^4k_3\ \mathrm{Tr} \bigl[ C(t)
\Phi_B(k_1) \Phi_{D_s^*}(k_2) \Phi_K(k_3) H(k_1,k_2,k_3, t)
\bigr], \label{eq:convolution1}
\end{equation}
where $k_i$'s are momenta of light quarks included in each meson.  $C(t)$ is
Wilson coefficient which results from the radiative corrections at
short distance.  $\Phi_M$ is the wave function which describes
hadronization of the quark and anti-quark into the meson $M$. $H$
describes the four quark operator and the quark pair from the sea
connected by a hard gluon whose scale is at the order of $M_B$, so
the hard part $H$ can be perturbatively calculated.

Some  analytic formulas for the   decay amplitudes of $B  \to D_s^* K$ decays
   will be given in the next section.
  In section \ref{sc:neval}, we give the numerical results and discussion.
Finally, we conclude this study in section \ref{sc:concl}.

\section{Analytic formulas}\label{sc:formula}

We consider the $B$ meson at rest for simplicity. In the
light-cone coordinate, the $B$ meson momentum $P_1$, the $D_s^*$
meson momentum $P_2$ and $K$ meson momentum $P_3$ are taken to be:
\begin{equation}
       P_1 = \frac{M_B}{\sqrt{2}} (1,1,{\bf 0}_T), P_2 =
       \frac{M_B}{\sqrt{2}} (1,r^2,{\bf 0}_T), P_3 =
       \frac{M_B}{\sqrt{2}} (0,1-r^2,{\bf 0}_T) , \label{eq:momentun1}
\end{equation}
where $r= M_{D_s^*}/M_B$ and we neglect the $K$ meson's mass $M_K$.
 The $D_s^*$ meson's longitudinal  polarization
 vector $\epsilon$  is   given by
$ \epsilon= \frac{M_B}{\sqrt{2} M_{D_s^*}} (1,-r^2,{\bf 0}_T) $.
Denoting the light (anti-)quark momenta in $B$, $D_s^*$ and $K$
mesons as $k_1$, $k_2$, and $k_3$, respectively, we can choose
$
k_1 = (x_1P_1^+,0,{\bf k}_{1T})$, $k_2 = (x_2 P_2^+,0,{\bf
k}_{2T})$, $
 k_3 = (0, x_3 P_3^-,{\bf k}_{3T})$.
The decay amplitude in eq.(\ref{eq:convolution1})
leads to:
\begin{multline}
 \mbox{Amplitude}
\sim \int\!\!
d x_1 d x_2 d x_3
b_1 d b_1 b_2 d b_2 b_3 d b_3 \\
\mathrm{Tr} \bigl[ C(t) \Phi_B(x_1,b_1)
\Phi_{D_s^*}(x_2,b_2,\epsilon) \Phi_K(x_3, b_3) H(x_i, b_i,\epsilon,
t) S_t(x_i)\, e^{-S(t)} \bigr], \label{eq:convolution2}
\end{multline}
where $b_i$ is the conjugate space coordinate of $k_{iT}$, and $t$
is the largest energy scale in $H$.
The large logarithms $\ln( m_W/t)$ coming from
QCD radiative corrections to four quark operators are included in
the Wilson coefficients $C(t)$.
The last term,
$e^{-S(t)}$, contains two kinds of logarithms. One of the large
logarithms is due to the renormalization of ultra-violet
divergence $\ln tb$, the other is double logarithm $\ln^2 b$ from
the overlap of collinear and soft gluon corrections. This Sudakov
form factor suppresses the soft dynamics effectively \cite{soft}.
Thus it makes perturbative calculation of the hard part $H$
applicable at intermediate scale, i.e.$M_B$ scale.

The $B$ meson wave function for incoming state and the $D_s^*$ and
$K$ meson wave function for outgoing state with up to twist-3
terms are written as:
\begin{equation}
 \Phi_{B,\alpha\beta}(x,b) = \frac{i}{\sqrt{2N_c}}
\left[ (\not \! P_1 \gamma_5)_{\alpha\beta} + M_B
\gamma_{5\alpha\beta} \right]  \phi_B(x,b),
\end{equation}
\begin{equation}
 \Phi_{D_s^*,\alpha\beta}(x,b) = \frac{i}{\sqrt{2N_c}}
\left[  \not \! \epsilon (\not \! P_2+ M_{D_s^*})_{\alpha\beta}
\right] \phi_{D_s^*}(x,b),
\end{equation}
\begin{align}
 \Phi_{K,\alpha\beta}(x_3,b_3) &= \frac{i}{\sqrt{2N_c}} \Bigl[ \gamma_5 \not \! P_3
\phi_K^A(x_3,b_3)
+ m_{0K} \gamma_5 \phi_K^P(x_3,b_3) \nonumber \\
& \qquad\qquad\qquad\qquad\qquad
 + m_{0K} \gamma_5 (\not v \not n - 1)\phi_K^T(x_3,b_3)
\Bigr]_{\alpha\beta},
\end{align}
 where $v = (0,1,{\bf 0}_T ) \propto P_3$, $n = (1,0,{\bf 0}_T)
 $, and $m_{0K} = M_K^2/(m_u + m_s)$.

Since the wave functions are process independent, one can use the
same forms constraint by other decay channels \cite{PQCD} to make predictions
here.
Now the hard part $H$ in eq.(\ref{eq:convolution2}) is the only channel
dependent part for us to  calculate perturbatively.


\subsection{$B^0 \to D_s^{*-} K^+$ decay}

\begin{figure}[htbp]
   \begin{center}
     \epsfig{file=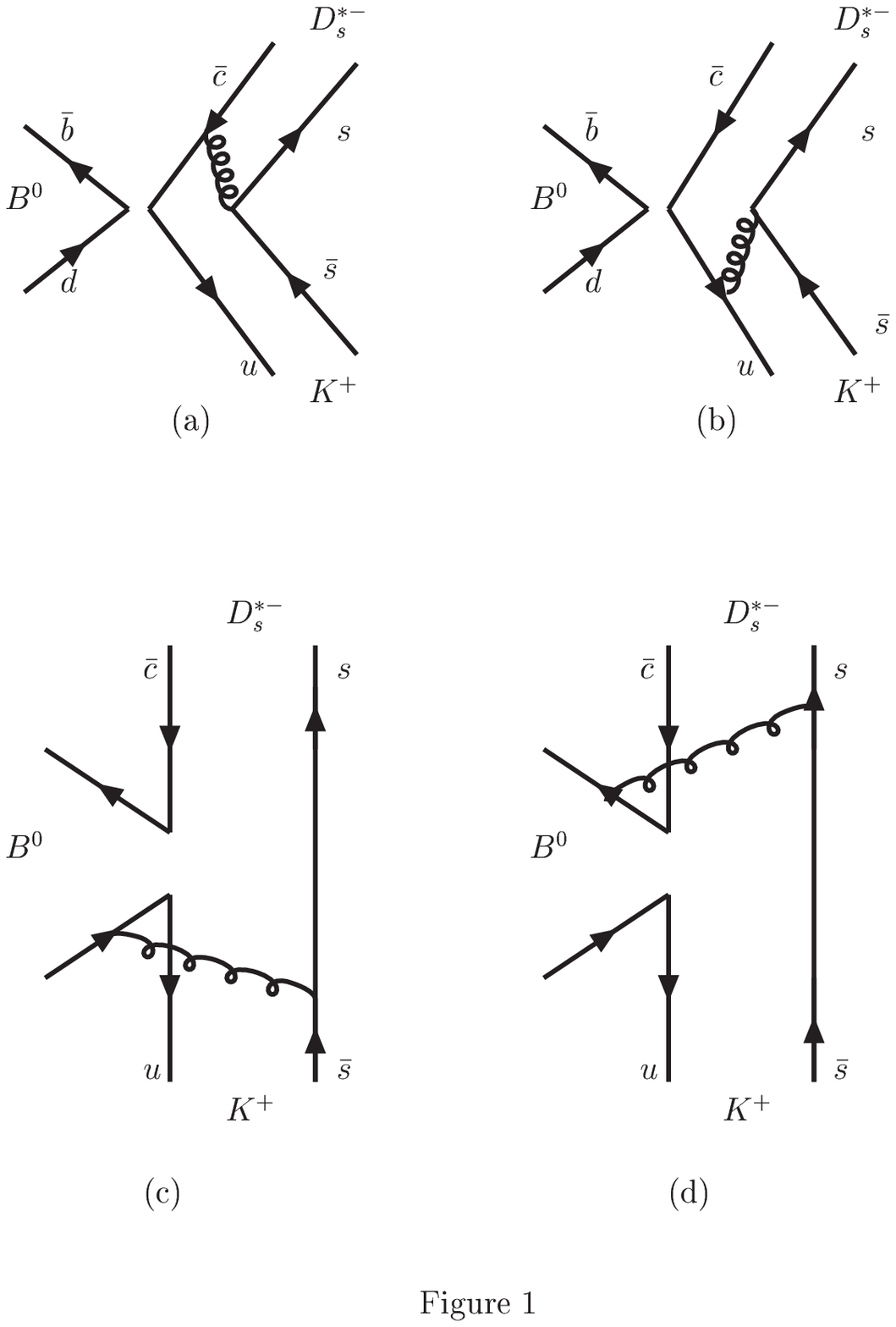,bbllx=5.8cm,bblly=10.4cm,bburx=11cm,bbury=23cm,%
 width=3cm,angle=0}
   \end{center}
   \caption{Diagrams for $B^0 \to D_s^{*-} K^+$ decay. The factorizable
   diagrams (a),(b) contribute to $F_a^{(2)}$, and nonfactorizable (c),
   (d) do to $M_a$.}
   \label{fig:diagrams1}
  \end{figure}

In the decay $B^0 \to D_s^{*-} K^+$, the effective Hamiltonian at the
scale lower than $M_W$   is given as
\cite{Buchalla:1996vs}
\begin{gather}
 H_\mathrm{eff} = \frac{G_F}{\sqrt{2}} V_{cb}^*V_{ud} \left[
C_1(\mu) O_1(\mu) + C_2(\mu) O_2(\mu) \right], \\
  O_1 = (\bar{b}d)_{V-A} (\bar{u}c)_{V-A}, \quad
 O_2 = (\bar{b}c)_{V-A} (\bar{u}d)_{V-A},
\end{gather}
where $C_{1,2}(\mu)$ are Wilson coefficients at renormalization
scale $\mu$, and summation in $\mathrm{SU}(3)_c$ color's index
$\alpha$ and chiral projection, $\sum_\alpha \bar{q}_\alpha
\gamma^\nu(1-\gamma_5)q'_\alpha$ are abbreviated to
$(\bar{q}q')_{V-A}$. The lowest order diagrams of $B^0 \to
D_s^{*-} K^+$ are drawn in Fig.\ref{fig:diagrams1} according to
this effective Hamiltonian.
There are altogether four quark diagrams. Just as what we said above, the decay $B \to
D_s^* K$ only has annihilation diagrams.

By calculating the hard part $H$ at the first order of $\alpha_s$,
we get the following analytic formulas. With the meson wave
functions, the amplitude for the factorizable annihilation
diagrams in Fig.\ref{fig:diagrams1}(a) and (b) results in
$F_a^{(i=2)}$ as:
\begin{multline}
F_a^{(i)} = -16\pi C_F M_B^2 \int_0^1 \!\! dx_2 dx_3
 \int_0^\infty  \!\!  b_2 db_2\, b_3 db_3\ \phi_{D_s^*}(x_2,b_2) \\
\times \Bigl[ \bigl\{
(1-r^2)\left( 1 - (1-r^2)x_3 \right) \phi_K^A(x_3,b_3)
+ r (1 - r^2)x_3 r_K \phi_K^P(x_3,b_3) \\
- r (1+r^2) r_K \phi_K^T(x_3,b_3)
\bigr\} E_{f}^i(t_a^1) h_a(x_2,x_3,b_2,b_3)
- \bigl\{
(1-r^2)x_2\phi_K^A(x_3,b_3) \\
+ 2r (1-r^2-x_2) r_K \phi_K^P(x_3,b_3) \bigr\} E_{f}^i(t_a^2)
h_a(1-x_3,1-x_2,b_3,b_2) \Bigr]. \label{eq:Fa}
\end{multline}
In the function, $C_F = 4/3$ is the group factor of
$\mathrm{SU}(3)_c$, and $r_K = m_{0K}/M_B$. The
functions $E_{f}^i$, $t_a^{1,2}$, $h_a$ will be given in the
appendix. The  distribution amplitudes $\phi_M$, are given in the next
section. The amplitude for the nonfactorizable annihilation
diagrams in Fig.\ref{fig:diagrams1}(c) and (d) results in
\begin{multline}
M_a  =  \frac{1}{\sqrt{2N_c}} 64\pi C_F M_B^2
\int_0^1 \!\! dx_1 dx_2 dx_3
 \int_0^\infty \!\! b_1 db_1\, b_2 db_2\
\phi_B(x_1,b_1) \phi_{D_s^*}(x_2,b_2) \\
\times \Bigl[
\bigl\{ (1-r^2) \left((1-r^2)(1 - x_3) - r^2 x_2 \right) \phi_K^A(x_3,b_2) \\
 - r \left(x_2 - (1-r^2)(1-x_3) \right) r_K \phi_K^P(x_3,b_2) \\
 - r \left(x_2 + (1-r^2)(1-x_3) \right) r_K \phi_K^T(x_3,b_2)
\bigr\}
E_{m}(t_{m}^1) h_a^{(1)}(x_1, x_2,x_3,b_1,b_2) \\
- \bigl\{
(1-r^2) \left( (1-r^2)x_2 + r^2 \right) \phi_K^A(x_3,b_2) \\
 + r (x_2 - (1-r^2)(1-x_3)) r_K \phi_K^P(x_3,b_2) \\
 + r (x_2-2r^2-(1-r^2)(1+x_3)) r_K \phi_K^T(x_3,b_2)
\bigr\}
E_{m}(t_{m}^2) h_a^{(2)}(x_1, x_2,x_3,b_1,b_2)
\Bigr],
\label{eq:Ma1}
\end{multline}
where $x_1$ dependence in the numerators of the hard part are
neglected by the assumption $x_1 \ll x_2, x_3$.

Comparing with $B^0 \to D_s^- K^+$ \cite{lu:bdsk}, we find that
the leading twist contribution, which is proportional to $\phi_K^A$, is almost the same.
However the subleading twist contribution, which is proportional
to $r~r_K$, change significantly, especially for the
 $\phi_K^P(x_3,b_2)$ terms in (\ref{eq:Fa}) and (\ref{eq:Ma1}).

The total decay amplitude for $B^0 \to D_s^{*-} K^+$ decay is
given as
$
 A = f_B F_a^{(2)}  + M_a
$.
The decay width is then
\begin{equation}
 \Gamma(B^0 \to D_s^{*-} K^+) = \frac{G_F^2 M_B^3}{128\pi} (1-r^2)
|V_{cb}^*V_{ud} A|^2.
\label{eq:neut_width}
\end{equation}
The charge conjugate decay  $\overline{B}^0 \to D_s^{*+} K^-$ is the same value as
$B^0 \to D_s^{*-} K^+$ just because $|V_{cb}^*V_{ud}|$ is the same
as
$|V_{cb}V_{ud}^*|$. Since there is only one kind of CKM phase
involved in the decay, there is no CP violation in the standard
model  for this decay channel.

\subsection{$B^+ \to D_s^{*+} \overline{K}^0$ decay}

The effective Hamiltonian related to $B^+ \to D_s^{*+}
\overline{K}^0$ decay is given as  \cite{Buchalla:1996vs}
\begin{gather}
 H_\mathrm{eff} = \frac{G_F}{\sqrt{2}} V_{ub}^*V_{cd} \left[
C_1(\mu) O_1(\mu) + C_2(\mu) O_2(\mu) \right], \\
  O_1 = (\bar{b}d)_{V-A} (\bar{c}u)_{V-A}, \quad
 O_2 = (\bar{b}u)_{V-A} (\bar{c}d)_{V-A}.
\end{gather}
The amplitude for the factorizable annihilation diagrams results
in $-F_a^{(i=1)}$. The amplitude for the nonfactorizable
annihilation diagrams is given as
\begin{multline}
M_a^{\prime} =  \frac{1}{\sqrt{2N_c}} 64\pi C_F M_B^2
\int_0^1\!\!\! dx_1 dx_2 dx_3
 \int_0^\infty\!\!\!\!\! b_1 db_1\, b_2 db_2\
 \phi_B(x_1,b_1) \phi_{D_s^*}(x_2,b_2) \\
\times \Bigl[ \bigl\{ (1-2r^2)x_2 \phi_K^A(x_3,b_2)
 + r \left(x_2-(1-r^2)(1-x_3)\right) r_K \phi_K^P(x_3,b_2) \\
 + r \left(x_2+(1-r^2)(1-x_3)\right) r_K \phi_K^T(x_3,b_2)
\bigr\}
E_{m}'(t_{m}^1) h_a^{(1)}(x_1, x_2,x_3,b_1,b_2) \\
- \bigl\{
(1-r^2) \left(1-x_2r^2-(1-r^2)x_3 \right) \phi_K^A(x_3,b_2) \\
 - r \left( x_2 - (1-r^2)(1-x_3) \right) r_K \phi_K^P(x_3,b_2) \\
 - r \left(x_2 -2r^2-(1-r^2)(1-x_3) \right) r_K \phi_K^T(x_3,b_2)
\bigr\}E_{m}'(t_{m}^2) h_a^{(2)}(x_1, x_2,x_3,b_1,b_2)
\Bigr].
\label{eq:Ma2}
\end{multline}
Comparing with the decay to two pseudo-scaler final states
$B^+ \to D_s^+ \overline{K}^0$ \cite{lu:bdsk}, the change only
occur in $r^2$ terms. Therefore, the non-factorizable contribution
in $B^+$ decay does not change much.

The total decay amplitude $A'$ and decay width $\Gamma$ for $B^+
\to D_s^{*+} \overline{K}^0$ decay are given as
\begin{gather}
  A' = - f_B F_a^{(1)} + M_a^{\prime},
\label{eq:chrg_amp} \\
 \Gamma(B^+ \to D_s^{*+} \overline{K}^0) = \frac{G_F^2 M_B^3}{128\pi} (1-r^2)
|V_{ub}^*V_{cd} A'|^2 .
\label{eq:chrg_width}
\end{gather}
The decay width for CP conjugated mode, $B^- \to D_s^{*-} K^0$, is
the same value as $B^+ \to D_s^{*+} \overline{K}^0$. Similar to
the $B^0$ decay, there is also no CP violation in this decay
within standard model.

\section{Numerical evaluation}\label{sc:neval}

We use the wave function of $B$  meson written as  \cite{PQCD}
\begin{equation}
\phi_B(x,b) = N_B x^2(1-x)^2 \exp \left[
-\frac{M_B^2\ x^2}{2 \omega_b^2} -\frac{1}{2} (\omega_b b)^2
\right],
\end{equation}
  where
$N_B$ is a normalization factor.  Because the
mass difference between $D_s$ and $D_s^*$ is not large, for
simplicity, their wave functions are chosen to be the same \cite{PQCD,lu:bdsk}:
\begin{equation}
\phi_{D_s^*}(x,b) = \frac{3}{\sqrt{2 N_c}} f_{D_s^*} x(1-x)\{ 1 +
a_{D_s^*} (1 -2x) \}\exp \left[-\frac{1}{2} (\omega_{D_s^*} b)^2
\right].
\end{equation}
Since $c$ quark is much heavier than $s$ quark, this function is
peaked at $c$ quark side, i.e. small $x$ region.

The  $K$ meson wave functions  are given as
\begin{eqnarray}
\phi_K^A(x) &=& \frac{f_K}{2\sqrt{2 N_c}} 6 x(1-x)
\left\{ 1 - a_1^K \cdot 3 \xi +
a_2^K \cdot \frac{3}{2} (-1 + 5 \xi^2) \right\}, \\
\phi_K^P(x) &=& \frac{f_K}{2\sqrt{2 N_c}}
\left\{ 1 + a_{p1}^K \cdot \frac{1}{2} (-1 + 3\xi^2)
+ a_{p2}^K \cdot \frac{1}{8} (3 - 30\xi^2 +35\xi^4) \right\}, \\
\phi_K^T(x) &=& \frac{f_K}{2\sqrt{2 N_c}} (1-2x)
\left\{ 1 + a_T^K \cdot 3(-3 + 5 \xi^2) \right\},
\end{eqnarray}
where $\xi = 2 x -1$. The parameters of these distribution amplitudes
calculated from QCD sum rule \cite{Ball:1998je}  are given
 as
\begin{equation}
a_1^K = 0.17,\quad  a_2^K = 0.2,\quad  a_{p1}^K = 0.212,
 \quad a_{p2}^K = -0.148,\quad a_T^K = 0.0527,
\label{eq:parm_phiK}
\end{equation}
for $m_{0K} = 1.6 \mbox{ GeV}$.
In addition, we use the following input parameters \cite{lu:bdsk}:
\begin{gather}
f_B = 190 \mbox{ MeV},\  f_K = 160 \mbox{ MeV},\
f_{D_s^*} = 241 \mbox{ MeV}, \\
m_{0K} = 1.6 \mbox{ GeV},\ \omega_b = 0.4 \mbox{ GeV},\  a_{D_s^*}
= 0.3,\ \omega_{D_s^*} = 0.3 \mbox{ GeV}. \label{eq:shapewv}
\end{gather}
For branching ratio estimation, we use the   CKM matrix elements and the
lifetimes of B mesons as following \cite{pdg},
\begin{gather}
|V_{ud}|=0.9734, \  |V_{ub}|= 3.6 \times 10^{-3}, \\
|V_{cb}|= 41.2\times 10^{-3}, \  |V_{cd}|=0.224,
\label{eq:KMmatrix} \\
 \tau_{B^\pm}=1.67\times 10^{-12}\mbox{ s,}\quad
 \tau_{B^0}=1.54\times 10^{-12}\mbox{ s}.
\end{gather}
The predicted branching ratios are
\begin{gather}
 \mathrm{Br}(B^0 \to D_s^{*-} K^+) = 2.7 \times 10^{-5},
\label{eq:br1}\\
 \mathrm{Br}(B^+ \to D_s^{*+} \overline{K}^0) = 4.0 \times 10^{-8}.
\label{eq:br2}
\end{gather}
The branching ratio of $B^+ \to D_s^{*+} \overline{K}^0$ is much
smaller than that of $B^0 \to D_s^{*-} K^+$, due to the
suppression from CKM matrix element $\mid V_{ub}^* V_{cd} \mid << |V_{cb}^* V_{ud}|$.
C-H Chen has also given the branching ratio of $B^0 \to D_s^{*-}
K^+$ \cite{ch}, our result agrees with his.

In these processes of $B \to D_s^* K$,
 only the   longitudinal polarization  of $D_s^*$  has
contribution.
If we ignore the  difference between $f_{D_s}$ and $f_{D_s^*}$ and
the difference between $M_{D_s}$ and $M_{D_s^*}$, the branch
ratios are thought to be the same as the corresponding $B\to D_s K$ decays.
But in fact the branching ratio of
$B^0 \to D_s^{*-} K^+$ is a little smaller than that of $B^0 \to D_s^-
K^+$,
because the contribution of twist-3 wave function in $B^0 \to
D_s^{*-} K^+$ decay become negative. Although the larger $f_{D_s^*}$ makes the branching ratio
larger,  a larger $M_{D_s^*}$ leads to a smaller branching ratio. The effect of
$M_{D_s^*}$ is more dominant than that of $f_{D_s^*}$.
In
decays of charged $B^+$, $f_{D_s^*}$ makes branching ratios larger, and
$M_{D_s^*}$ has little effect, such that the $B^+\to D_s^{*+} \bar K^0$ decay
in this case has a little
larger branching ratio than $B^+\to D_s^+ \bar K^0$.

For the experimental side,  there are only upper limits given at $90$\%
confidence level \cite{pdg},
\begin{align}
 \mathrm{Br}(B^0 \to D_s^{*-} K^+)
& < 1.7 \times 10^{-4},\label{eq:brex1}\\
 \mathrm{Br}(B^+ \to D_s^{*+} \overline{K}^0)
& < 1.1 \times 10^{-3}.\label{eq:brex2}
\end{align}
Obviously, our results are consistent with the data.

In addition to the perturbative annihilation contributions, there
is also a hadronic picture for the $B^0 \to D_s^{*-} K^+$.  The
$B$ meson decays into $D^-$ and $\rho^+$, the secondary
particles then exchanging a $K^0$, scatter into $D_s^{*-}$, $ K^+$ through final state
interaction afterwards.
For charged decay, $B^+$ meson decays into $D^0\rho^+$ then scatter into
$D_s^{*+}$ and $\bar K^0$ by exchanging a $K^+$.
 But this picture can not be calculated
accurately. In \cite{lu:bdsk}, the results from PQCD approach for $B^0 \to D_s^-K^+$ decay
 were consistent with
the experiment, which shows that the soft final state interaction
may not be important.

The calculated branching ratios in PQCD are   sensitive to various parameters, such
as parameters in eqs.(\ref{eq:parm_phiK}-\ref{eq:shapewv}).
It is necessary to give the sensitivity  of the branching ratios when we
choose the parameters to some extent.
Table~\ref{tb:sensitivity} shows the sensitivity of the branch
ratios to $30$\% change of $a_{D_s^*}$,
 $\omega_{D_s^*}$ and $\omega_b$.
Here we don't present the
sensitivity to $a_{p1,p2,T}^K$ because the branching ratios are
insensitive to them.
It is found that the uncertainty of the predictions on PQCD is mainly due to
 $ \omega_{D_s^*}$, which characterizes the shape of $D_s^*$ wave function.

 \begin{table}[htbp]
\caption{The sensitivity of the decay branching ratios to $30$\% change of $
\omega_{D_s^*}$,  $a_{D_s^*}$ and $\omega_b$.} \label{tb:sensitivity}
\begin{center}
\begin{tabular}[t]{r|cc}
 \hline     \hline
& $(10^{-5})$ & $(10^{-8})$ \\
 $ \omega_{D_s^*}$ & $\mathrm{Br}(B^0 \to D_s^{*-} K^+)$ &
$\mathrm{Br}(B^+ \to D_s^{*+} \overline{K}^0)$ \\
 \hline
 $0.21$ & $3.17$ & $4.66$ \\
 $0.30$ & $2.65$ & $3.95$ \\
 $0.39$ & $2.11$ & $3.20$ \\
 \hline
 \hline
 $a_{D_s^*}$ & $\mathrm{Br}(B^0 \to D_s^{*-} K^+)$ &
$\mathrm{Br}(B^+ \to D_s^{*+} \overline{K}^0)$ \\
 \hline
 $0.21$ & $2.40$ & $3.68$ \\
 $0.30$ & $2.65$ & $3.95$ \\
 $0.39$ & $2.92$ & $4.24$ \\
 \hline
 \hline
 $\omega_b$ & $\mathrm{Br}(B^0 \to D_s^{*-} K^+)$ &
$\mathrm{Br}(B^+ \to D_s^{*+} \overline{K}^0)$ \\
 \hline
 $0.21$ & $3.12$ & $3.87$ \\
 $0.40$ & $2.65$ & $3.95$ \\
 $0.52$ & $2.31$ & $4.18$ \\
 \hline
\end{tabular}
\end{center}
\end{table}

 From the above discussions, we can derive the uncertainties of the branching ratios
 within the suitable
ranges on $ \omega_{D_s^*}$, $a_{D_s^*}$ and $\omega_b$. The
branching ratios normalized by the decay constants and the CKM matrix
elements result in
\begin{gather}
\mathrm{Br}(B^0 \to D_s^{*-} K^+) = (2.7\pm 0.6) \times
10^{-5} \left( \frac{f_B\ f_{D_s^*}}{190\mbox{ MeV}\cdot 240\mbox{
MeV}} \right)^2 \left( \frac{|V_{cb}^*\ V_{ud}|} {0.0412\cdot
0.9734} \right)^2
, \\
\mathrm{Br}(B^+ \to D_s^{*+} \overline{K}^0) =
(4.0 \pm 0.8) \times 10^{-8} \left( \frac{f_B\
f_{D_s^*}}{190\mbox{ MeV}\cdot 240\mbox{ MeV}} \right)^2 \left(
\frac{|V_{ub}^*\ V_{cd}|} { 0.0036 \cdot 0.224} \right)^2.
\end{gather}


\section{Conclusion}\label{sc:concl}

In this paper, we calculate the $B^0 \to D_s^{*-} K^+$ and $B^+
\to D_s^{*+} \overline{K}^0$ decays in PQCD approach. These two
decays occur purely via annihilation type diagrams because the
four quarks in final states are not the same as  the ones in $B$
meson.
 We argue that the soft final state interaction may be small in B
 decays. The PQCD calculation of annihilation decays is reliable.
  The branching ratio for $B^0 \to D_s^{*-}K^+$ decay is sizable at order of
  $10^{-5}$,
which can be measured
in the current $B$ factories Belle and BABAR.
The small branching ratio of $B^+ \to D_s^{*+} \bar K^0$ ($10^{-8}$) predicted in the SM,
makes it
sensitive to new physics contributions, which may be studied in the future   LHC-B experiment.

 \section*{Acknowledgments}

 This work is partly supported by National Science Foundation of
 China under Grant (No. 90103013 and 10135060).

\begin{appendix}

\section{Some formulas used in the text}

The function $E_f^i$, $E_m$, and $E'_m$
including Wilson coefficients
are defined as
\begin{gather}
 E_{f}^i(t) = a_i(t) \alpha_s(t)\, e^{-S_D(t)-S_K(t)}, \\
 E_{m}(t) = C_2(t) \alpha_s(t)\, e^{-S_B(t)-S_D(t)-S_K(t)}, \\
 E_{m}'(t) = C_1(t) \alpha_s(t)\, e^{-S_B(t)-S_D(t)-S_K(t)},
\end{gather}
 where
\begin{equation}
 a_1(t) =  {C_1(t)}/{3} + C_2(t),\quad
 a_2(t) = C_1(t) +  {C_2(t)}/3,
\end{equation}
and $S_B$, $S_D$, and $S_K$ result from summing both double logarithms
caused by soft gluon corrections and single ones due to
the renormalization of ultra-violet divergence.
Those factors are
 given in ref.
\cite{PQCD,lu:bdsk}.
   In
 the numerical analysis we use
  leading logarithms expressions for Wilson coefficients
 $C_{1,2}$ presented in ref.\cite{Buchalla:1996vs,luy}.

The functions $h_a$, $h_a^{(1)}$, and
$h_a^{(2)}$  in the decay amplitudes consist of two parts: one is
the jet function $S_t(x_i)$ derived by the threshold
resummation \cite{L3}, the other is the propagator of virtual quark
and gluon. They are defined by
\begin{align}
& h_a(x_2,x_3,b_2,b_3) = S_t(1-x_3)\left( \frac{\pi i}{2}\right)^2
H_0^{(1)}(M_B\sqrt{(1-r^2)x_2(1-x_3)}\, b_2) \nonumber \\
&\times \left\{
H_0^{(1)}(M_B\sqrt{(1-r^2)(1-x_3)}\, b_2)
J_0(M_B\sqrt{(1-r^2)(1-x_3)}\, b_3)
\theta(b_2 - b_3) + (b_2 \leftrightarrow b_3 ) \right\},
\label{eq:propagator1} \\
& h^{(j)}_a(x_1,x_2,x_3,b_1,b_2) = \nonumber \\
& \biggl\{
\frac{\pi i}{2} \mathrm{H}_0^{(1)}(M_B\sqrt{(1-r^2)x_2(1-x_3)}\, b_1)
 \mathrm{J}_0(M_B\sqrt{(1-r^2)x_2(1-x_3)}\, b_2) \theta(b_1-b_2)
\nonumber \\
& \qquad\qquad\qquad\qquad + (b_1 \leftrightarrow b_2) \biggr\}
 \times\left(
\begin{matrix}
 \mathrm{K}_0(M_B F_{(j)} b_1), & \text{for}\quad F^2_{(j)}>0 \\
 \frac{\pi i}{2} \mathrm{H}_0^{(1)}(M_B\sqrt{|F^2_{(j)}|}\ b_1), &
 \text{for}\quad F^2_{(j)}<0
\end{matrix}\right),
\label{eq:propagator2}
\end{align}
where $\mathrm{H}_0^{(1)}(z) = \mathrm{J}_0(z) + i\, \mathrm{Y}_0(z)$, and
$F_{(j)}$s are defined by
\begin{equation}
 F^2_{(1)} = (1-r^2)(x_1 -x_2)(1- x_3),\
F^2_{(2)} = x_1 +x_2+(1-r^2)(1-x_1-x_2)(1-x_3).
\end{equation}
We adopt the parametrization for $S_t(x)$ of the factorizable
contributions,
\begin{equation}
 S_t(x) = \frac{2^{1+2c}\Gamma(3/2 +c)}{\sqrt{\pi} \Gamma(1+c)}
[x(1-x)]^c,\quad c = 0.3,
\end{equation}
which is proposed in ref.~\cite{Kurimoto:2001zj}. The hard scale
$t$'s in the amplitudes are taken as the largest energy scale in
the hard part $H$ to kill the large logarithmic radiative corrections:
\begin{gather}
 t_a^1 = \mathrm{max}(M_B \sqrt{(1-r^2)(1-x_3)},1/b_2,1/b_3), \\
 t_a^2 = \mathrm{max}(M_B \sqrt{(1-r^2)x_2},1/b_2,1/b_3), \\
 t_{m}^j = \mathrm{max}(M_B \sqrt{|F^2_{(j)}|},
M_B \sqrt{(1-r^2)x_2(1-x_3) }, 1/b_1,1/b_2).
\end{gather}

\end{appendix}

\begin{thebibliography}{99}

\bibitem{pqcd}G.P. Lepage and S. Brosky, Phys. Rev. D22, 2157
(1980);
J. Botts and G. Sterman, Nucl. Phys. B225, 62 (1989).

\bibitem{luy}C.-D. L\"u, K. Ukai and M.-Z. Yang,
Phys. Rev. D63, 074009 (2001).

\bibitem{kls} Y.-Y. Keum, H.-n. Li and A. I. Sanda,
Phys. Lett. B504, 6 (2001); Phys. Rev. D63, 054008 (2001).
\bibitem{ly}C.-D. L\"u and M.Z. Yang, Eur. Phys. J. C23, 275
(2002).

\bibitem{PQCD}
H.-n. Li, Phys. Rev. D64, 014019 (2001);
S. Mishima, Phys. Lett. B521, 252 (2001);
E. Kou and A.I. Sanda, Phys. Lett. B525, 240 (2002);
C.-H. Chen, Y.-Y. Keum, and H.-n. Li, Phys. Rev. D64, 112002 (2001);
A.I. Sanda and K. Ukai, Prog. Theor. Phys. 107, 421 (2002);
C.-H. Chen, Y.-Y. Keum, and H.-n. Li, Phys. Rev. D66, 054013 (2002);
M. Nagashima and H.-n. Li, hep-ph/0202127;
Y.-Y. Keum, hep-ph/0209002; hep-ph/0209208(to appear in PRL); hep-ph/0210127;
Y.-Y. Keum and A. I. Sanda, Phys. Rev. D67, 054009 (2003);
C.D. L\"u, M.Z. Yang, hep-ph/0212373, to appear in Eur. Phys. J. C.

\bibitem{lu:bdsk} K. Ukai, 287-290, Preceedings for 4th International Conference on B Physics and
                 CP Violation (BCP4), Japan, 2001;
 C.-D. L\"{u} and K. Ukai, hep-ph/0210206, to appear in Eur. Phys. J. C.
 \bibitem{ch} C.-H. Chen, hep-ph/0301154.
\bibitem{lu}C.D. L\"u, Eur. Phys. J. C24, 121 (2002).

\bibitem{bsw}M. Wirbel, B. Stech, M. Bauer, Z. Phys. C29, 637 (1985);
 M. Bauer, B. Stech, M. Wirbel, Z. Phys. C34, 103 (1987);
 A. Ali, G. Kramer and C.D. L\"u, Phys. Rev. D58, 094009
(1998); C.D. L\"u, Nucl. Phys. Proc. Suppl. 74, 227-230 (1999);
Y.-H. Chen, H.-Y. Cheng, B. Tseng, K.-C. Yang,
 Phys. Rev. D60, 094014 (1999).

\bibitem{bbns}M. Beneke, G. Buchalla, M. Neubert, C.T. Sachrajda,
Phys. Rev. Lett. 83, 1914 (1999); Nucl. Phys.
B591, 313 (2000).

\bibitem{soft} H.-n. Li and B. Tseng, Phys. Rev. D57, 443, (1998).

\bibitem{Buchalla:1996vs} G. Buchalla, A. J. Buras and
    M. E. Lautenbacher, Rev. Mod. Phys. 68, 1125 (1996).
 \bibitem{Ball:1998je} P. Ball, JHEP, 09, 005, (1998); JHEP, 01, 010, (1999).
\bibitem{pdg} Review of Particle Physics, K. Hagiwara {\it et al.},
Phys. Rev. D66, 010001 (2002).

 \bibitem{L3} H.-n. Li, hep-ph/0102013 (to appear in PRD).
\bibitem{Kurimoto:2001zj}T. Kurimoto, H.-n. Li, and A. I. Sanda,
Phys. Rev. D65, 014007 (2002).

\end{thebibliography}

\end{document}